\documentclass{sig-alt-mod}
\pdfoutput=1

\usepackage{times}
\usepackage[final]{hyperref}

\usepackage{ifthen}
\usepackage{xspace}
\usepackage{epsf}

\newtheorem{fact}{Fact}[section]

\newtheorem{theorem}[fact]{Theorem}

\newtheorem{proposition}[fact]{Proposition}

\newlength{\saveparindent}
\newlength{\saveparskip}
\newcommand{\sketch}[1]{
{\noindent {\it Proof Sketch.} {#1} \rule{2mm}{2mm} 
}
}

\newcommand{\Xomit}[1]{}
\newcommand{\omt}[1]{}
\newcommand{\supproof}[1]{}

\newcommand{\xhdr}[1]{\paragraph*{\bf #1}}

\def\halfrs{\vspace*{-3ex}}
\def\rs{}

\def\view{\phi}
\def\delay{\delta}
\def\rate{\rho}
\def\bb{{\cal H}}
\def\abb{{\cal H}^*}
\def\skel{{\cal G}}
\def\wgr{{\cal G}^{\delta}}
\def\fht{0.23}

\begin{document}

\conferenceinfo{KDD'08,} {August 24--27, 2008, Las Vegas, Nevada, USA.}
\CopyrightYear{2008}
\crdata{978-1-60558-193-4/08/08}


\title{The Structure of Information Pathways in a Social Communication Network}
\numberofauthors{3}
\author{
\alignauthor Gueorgi Kossinets \\
       \affaddr{Dept.\ of Sociology}\\
       \affaddr{Cornell University}\\       
       \affaddr{Ithaca, NY  14853}\\
       \affaddr{gk67@cornell.edu}\\
\alignauthor Jon Kleinberg \\
       \affaddr{Dept.\ of Computer Science}\\
       \affaddr{Cornell University}\\
       \affaddr{Ithaca, NY  14853}\\
       \affaddr{kleinber@cs.cornell.edu}\\
\alignauthor Duncan Watts \\
       \affaddr{Yahoo! Research}\\
       \affaddr{111 West 40th Street, 17th Fl.}\\
       \affaddr{New York, NY  10018}\\
       \affaddr{djw@yahoo-inc.com}\\[2ex]
}

\date{}

\maketitle

\newcommand{\UnnumberedFootnote}[1]{{\def\thefootnote{}\footnote{#1}
\addtocounter{footnote}{-2}}}

\begin{abstract} 
Social networks are of interest to researchers in part because they
are thought to mediate the flow of information in communities and 
organizations.  
Here we study the temporal dynamics of communication
using on-line data, including e-mail communication among the faculty
and staff of a large university over a two-year period.  We formulate
a temporal notion of ``distance'' in the underlying social network by
measuring the minimum time required for information to spread from one
node to another --- a concept that draws on the notion of 
vector-clocks from the study of distributed computing systems. 
We find that such temporal measures provide structural insights that are
not apparent from analyses of the pure social network topology.  
In particular, we define the {\em network backbone} to be the subgraph
consisting of edges on which information has the potential to flow the
quickest.  We find that the backbone is a sparse graph with a
concentration of both highly embedded edges and long-range bridges ---
a finding that sheds new light on the relationship between tie strength and
connectivity in social networks.  
\end{abstract}

\vspace{5mm}
\noindent
{\bf Categories and Subject Descriptors:}
H.2.8 {Database Management}: {Database Applications -- Data Mining}

\vspace{1mm}
\noindent
{\bf General Terms:} Measurement, Theory

\vspace{1mm}

\vspace{1mm}
\noindent
{\bf Acknowledgments:}
This research was supported in part by 
the Institute for Social and Economic Research and
Policy at Columbia University,
the Institute for the Social Sciences at Cornell University,
the James S. McDonnell Foundation,
the John D. and Catherine T. MacArthur Foundation,
a Google Research Grant, 
a Yahoo!~Research Alliance Grant,
and NSF grants
SES-0339023, CCF-0325453, IIS-0329064, CNS-0403340, and BCS-0537606.



\newpage

\vspace{1ex}
\section{Introduction} \label{sec:intro}

Large social networks serve as conduits for communication
and the flow of information
\cite{adamic-huberman-survey, granovetter-weak-ties};
but information only spreads on these networks as a result of discrete
communication events---such as e-mail or text messages, conversations,
meetings, or phone calls---that are distributed non-uniformly over
time \cite{gibson-scheduling,white-everyday-life}.  
Just because two individuals are
acquainted does not imply that they have communicated within some
particular time interval, in which case no information could have
passed directly between them.  Correspondingly, the indirect flow of
information between individuals requires a sequence of communication
events along a path of intermediaries linking them.  Although
straightforward to state, these observations pose additional
complications for the analysis of social networks,
and can have important
consequences for the relation between network structure and
information flow \cite{holme-reachability, onnela-phone-data}.

It has been a challenge to build reasonable models for 
the patterns of communication within a social network:
it is difficult to obtain data on social network structure
at a large scale, and more difficult still to obtain
complete data on the dynamics of 
a network's communication events over time.
Recent research working with such datasets has primarily studied communication
of an {\em event-driven} nature, looking at communication within 
a social network triggered by a particular event or activity;
such investigations have typically focused on communication
events that ripple through many nodes over
short time-scales following the triggering event.
Examples of this include cascades of e-mail recommendations
for products \cite{leskovec-ec06}, cascades of references
among bloggers \cite{adar-blogspace,gruhl-blogspace,leskovec-blogspace-sdm07},
the spread of e-mail chain letters \cite{liben-nowell-pnas08},
and the search for distant targets in a social network
\cite{dodds-swn-expmt,travers-swn}.

These types of event-driven communication, however, take place
against the backdrop of a much broader set of natural communication rhythms,
a kind of {\em systemic communication} that circulates information
continuously through the network.
Pairs of individuals communicate over time at very different rates,
for an enormous range of different reasons. 
Viewed cumulatively, this background
pattern enables information to piggy-back on everyday communication
and thus spread generally through the network.
This type of systemic communication has remained essentially
invisible in analyses of social networks over time, but
its properties arguably determine much about the rate at
which people in the network remain up-to-date on information
about each other.

\begin{figure}[t]
\begin{center}
\rs \rs 
\includegraphics[height=.25\textheight]{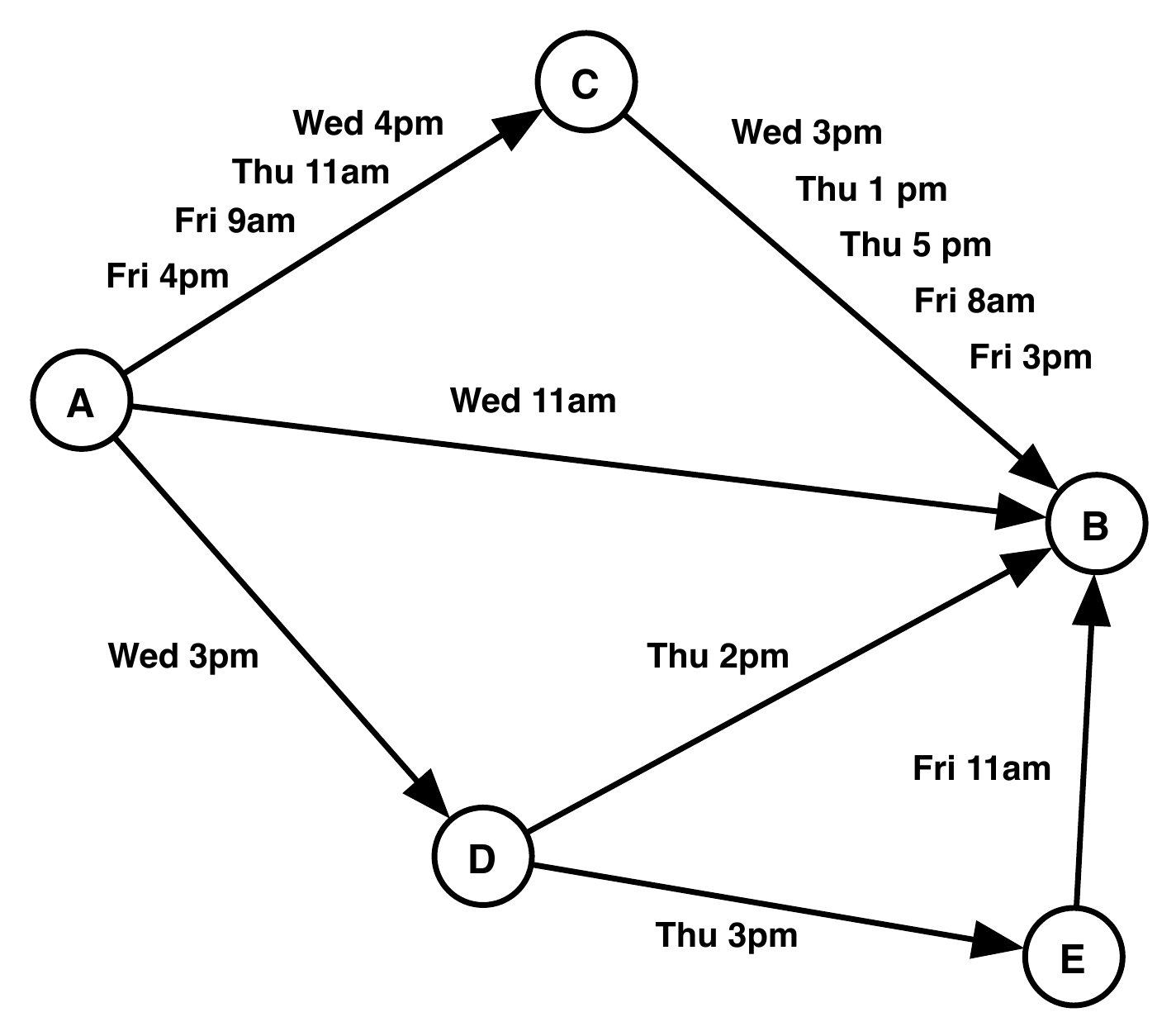}
\rs 
\caption{
Node $B$'s most recent potential information about node $A$
comes via node $C$, not directly from $A$.
\label{fig:timestamps}
}
\rs \rs 
\halfrs
\end{center}
\end{figure}

\xhdr{The present work: Systemic communication and information pathways}
We propose a framework for analyzing this kind of systemic communication,
based on inferring structural measures from the {\em potential} for
information to flow between different nodes.
To motivate this by an example, suppose we have the
complete communication history for a group of five people
over three days, as illustrated in Figure~\ref{fig:timestamps}.
(Edges are annotated with the one or more times at
which directed communication took place.)
For the sake of this example, let us assume that there
are no communication events outside the group that
are relevant to the analysis.
We can now ask questions such as the following: 
At 5pm on Friday, what is the most recent
information that node $B$ could possibly have about node $A$?
Clearly $B$ could have learned about $A$'s state as of Wednesday at 11am, when 
their last direct communication took place. However, further inspection
of the figure reveals that the most recent opportunity for information
to flow from $A$ to $B$ was in fact the Friday 
9am communication from $A$ to $C$, which was then
followed by the Friday 3pm    
communication from $C$ to $B$.

Without knowing anything about the content of the messages,
we will not necessarily know what, if anything, flowed between nodes,
but this sequence of timestamps gives us a global
picture of the {\em information pathways}, providing
the full set of potential conduits for information to flow
through the group of people.
From this structure, we can draw several conclusions.
First, still without knowing the message content,
we can conclude that anything that has happened to $A$
in the past eight hours will be unknown to $B$:
at Friday 5pm, $B$ is in a strong sense 
(at least) eight hours ``out-of-date'' with respect to $A$.
Second, assuming this interval of three days is typical of
the communication dynamics within this group of five people,
we can infer that direct communication does not generally provide $B$
with the strongest opportunities to learn information about $A$;
rather, the indirect $A$-$C$-$B$ path has the potential to 
transmit information from $A$ to $B$ much faster than the direct link.

We argue here that these latter two issues --- out-of-date
information and indirect paths --- are central to an understanding
of the patterns of systemic communication within a social network.
The notion of individuals being out-of-date with respect to
each other's information is an intuitively natural one,
and one finds implicit reflections of it in settings ranging from 
the study of physical systems to social processes and fictional narratives.
The physical world, for example, is governed by principle 
that we are at least $k$ years out-of-date with respect to any
point in space $k$ light-years away; the notion of the {\em light cone}
more generally characterizes the regions 
of space-time between which information 
can possibly have flowed \cite{taylor-relativity}.
In sociology, the premise that occasional encounters with 
distant acquaintances can provide important information
about new opportunities helps form the basis for
Granovetter's celebrated theory of the {\em strength of weak ties}
\cite{granovetter-weak-ties}.
And in yet a different direction, the idea that two 
individuals sometimes cannot know what has happened to one another,
over short time spans, arises as a literary device;
for example, in his novel {\em The Gift}, Vladimir Nabokov
provides the following grim but memorable image to convey
the idea that it took the character Yasha's family several
hours to learn of his suicide:
\begin{quote}
... no sooner had he reached her than
both of them heard the dull pop of the shot, while in Yasha's
room life went on for a few more hours as if nothing had happened ...
\cite{nabokov-gift}
\end{quote}

The role of indirect paths in social communication 
is also a crucial issue that
has received relatively little formal attention.
If we look at a social network represented simply as
an unweighted graph, then any time two nodes are joined
by an edge, this edge provides the most direct path between them.
If we have data, however, on the times or rates at which
communication actually takes place across edges, then
we can discover --- as in Figure~\ref{fig:timestamps} --- 
that often information has the potential to flow much 
more rapidly via multi-step paths.
In a sense, then, the $A$-$C$-$B$ path in Figure~\ref{fig:timestamps}
can be viewed as a ``triangle-inequality violation'', in 
that a two-step path can be faster than a one-step path.
One finds intuitively natural reflections of this principle
in everyday life: a manager who talks to each of two employees much
more frequently than they talk to each other, or a parent
who talks to each of two adult children much more frequently
than they talk to each other.
We will see later that the structure of communication in
real social networks is in fact dominated by such
violations of the triangle inequality.

\xhdr{The present work: Vector clocks and backbone structures}
We now proceed to study these notions of out-of-date information
and indirect paths using data for which
we have complete histories of communication events over
long periods of time.
Our main dataset is a complete set of anonymized e-mail logs
among all faculty and staff at a large university over two years
\cite{kossinets-email}.  
We will use this university e-mail dataset 
as the primary focus of discussion in the present work;
but at the end, we also discuss the results of our analyses
on two other sources of data: 
the Enron e-mail corpus \cite{klimt-enron}, a widely-used
dataset containing of e-mail communication among executives 
from the (now-defunct) Enron corporation;
and also, in a quite different domain, the complete set
of user-talk communications among admins and
high-volume editors on Wikipedia.
Taken together, these datasets 
thus represent a range of different settings in
which the patterns of systemic communication within a large group
are integral to the workflow of the group.
We find broadly similar patterns of results across all of them.

We analyze the issue of out-of-date information by
adapting ideas from the field of distributed computing,
which has also had to deal with the problem of potential
information flow among different computing hosts ---
determining, for example, which machines might be affected
if a given host is compromised at a given point in time.
In particular, we use the notion of {\em vector clocks}
introduced by Lamport and refined by Mattern to study how information
spreads in distributed systems \cite{lamport-clocks,mattern-clocks}.  
(Mattern's development, among other things, 
draws interesting analogies with notions of
simultaneity and light-cones from special relativity \cite{mattern-clocks}.)

Next, we formalize the notion of indirect paths by
defining the {\em network backbone} --- the subset 
of edges in the social network that are not 
bypassed by a faster alternate path.
We propose several related definitions of the backbone, and
for all formulations we find that the backbone is a very sparse
subgraph consisting of a mixture of highly embedded edges
and longer-range bridges.
Finally, we consider how potential 
information flow would be affected if communication
were sped up or slowed down on certain backbone edges,
and use this to draw conclusions about the effect
of local communication rates on the global circulation
of information.

In the end, it is important to reiterate, we are using
these notions of potential information flow to draw
structural conclusions about social communication networks
in their everyday operation.  We do not attempt to map
the actual contents of messages as they are being sent,
nor are we focusing on the effects of one-time, ``special''
events that can generate novel communication flows.
Rather, our goal is to approach a dual, and largely unstudied, issue ---
how everyday patterns of communication suggest certain temporal
notions of distance that are distinct from the picture 
that an unweighted graph provides, and how these patterns
cause certain sparse sets of pathways to emerge as
the lines along which information has the ability to
flow the quickest.

\xhdr{Further related work}
The complete traces of communication within a network of people
has been studied at moderate scales in recent years
\cite{adamic-how-search,eckmann-email}, and very recently there have
been analyses of very large-scale networks based on
phone calls \cite{onnela-phone-data} 
and instant messaging \cite{leskovec-msn-im}.
These studies, however, have focused on structural properties 
of the networks different from the definitions we propose here.
As noted above, a number of other recent lines of research have focused
on cascading communication triggered by specific events
\cite{
adar-blogspace,
dodds-swn-expmt,
gruhl-blogspace,
leskovec-ec06,
leskovec-blogspace-sdm07,
liben-nowell-pnas08},
but this work too addresses issues that are quite different
from our focus here.

The notion of a graph annotated with the times at which the
nodes communicated has been studied at a theoretical level
\cite{berman-temporal,cheng-temporal,holme-reachability,kempe-temporal}.
Holme has explored some of the theoretical definitions on 
network datasets \cite{holme-reachability}, though in different
directions from what we do here.

Finally, the sub-field of distributed computing concerned
with {\em epidemic} or {\em gossip-based} algorithms has
focused on designing communication patterns that spread
information quickly \cite{demers-epidemic}.
In contrast, we focus here on systems that are not designed,
but where analyses of the communication patterns over time
can nonetheless provide us with insights into 
underlying structures in the network.

\section{Vector Clocks and Latency}

The basic structure of the data we consider is as follows.
We are given a set $V$ of people (nodes) communicating over a time 
interval $[0,T]$, and we have a complete trace of the communication
events among them.  
Each recorded communication event consists of a triple $(v,w,t)$,
indicating that 
node $v$ sent a message to node $w$ at time $t$.
We also define an unweighted directed graph that simply represents 
the pairs who ever communicated; thus, we define the
{\em communication skeleton} $\skel$ to be the graph 
on $V$ with an edge $(v,w)$ if $v$ sent at least one message to $w$
during the observation period $[0,T]$.

We begin by briefly reviewing the approach of Lamport, Mattern,
and others in the line of distributed computing research
aimed at formalizing temporal lags between nodes in a network
\cite{lamport-clocks,mattern-clocks}.
To start, we consider a node $v$ at time $t$ and try to determine how 
``up-to-date'' its information about another node $u$ could be.
We can quantify this by asking the following question:
what is the largest $t' < t$ for which a piece of 
information originating at time $t'$ at $u$ could 
be transmitted through a sequence of communications and
still arrive at $v$ by time $t$?
We call this largest $t'$ the {\em view} that $v$ has of $u$
at time $t$, and denote it by $\view_{v,t}(u)$.
The amount by which $v$'s view of $u$ is ``out-of-date''
at time $t$ is given by $t - \view_{v,t}(u)$;
we will call this the {\em information latency} of $u$ with respect to $v$
at time $t$.
For example, in Figure~\ref{fig:timestamps}, the views that 
$B$ has of $A$, $C$, $D$, and $E$ at Friday 5pm are,
respectively, Fri 9am, Fri 3pm, Thu 3pm, and Fri 11am;
and hence the latencies are 8 hours, 2 hours, 26 hours, and 6 hours.
(We will define $\view_{v,t}(v) = t$ for all $v$ and $t$:
$v$ is always completely up-to-date with respect to itself.)
Finally, we can take all the views of other nodes that $v$ has at time $t$
and write it as a single
vector $\view_{v,t} = (\view_{v,t}(u) : u \in V)$.
We refer to $\view_{v,t}$ as the {\em vector clock} of
$v$ at time $t$ \cite{lamport-clocks,mattern-clocks}.

There is a simple and efficient algorithm to compute the
vector clocks for all nodes at all times in $[0,T]$
by a single pass through the history of communication events
$(v,w,t)$, ordered by increasing $t$ \cite{lamport-clocks,mattern-clocks}.
The algorithm proceeds by maintaining each vector clock $\view_v$
as a variable that is updated when $v$ receives a communication.
We initialize each vector $\view_v$ 
to have a special null symbol $\bot$ in each coordinate 
(except that $\view_{v,0}(v) = 0$),
indicating that no node has yet heard, even indirectly, from any other.
Then, in general, when we process event $(v,w,t)$, we update
the vector clock $\view_w$ to be the coordinatewise maximum
of the current values of $\view_v$ and $\view_w$ (treating
$\bot$ as smaller than any number); this reflects the fact
that when $v$ sends a message to $w$, node $w$ gets a view 
of each node that is the more recent of $v$'s view and $w$'s view.
(When we process this event, we set $\view_{v,t}(w) = t$,
since $v$ has just heard from $w$.)
We run this procedure for all events, thus obtaining a 
value for the vector clock of each node at each point in time.

\xhdr{Latencies in Social Network Data}
We now examine these latency measures in the context of real social
communication data.  Again, we focus on 
our university e-mail dataset,
but in the final section
we also discuss our other datasets --- the Enron corpus and
the communications among Wikipedia editors.

For the university e-mail study
we start from the complete set of communication events
among the 8160 faculty and staff at a large university over two years,
and then we preprocess this set in two ways.
First, it is an interesting open question to consider the appropriate
role for messages with large recipient lists in this type of analysis;
however, for the present study, we eliminate them by considering only
messages with at most $c$ recipients other than the sender,
for small values of $c$ (ranging between $1$ and $5$).\footnote{We use a 
heuristic based on timestamps and file sizes to detect 
multi-recipient messages that a mail client or server 
has serialized into many single-recipient messages for
purposes of transmission.}
Messages with a single recipient account for $82\%$ of all messages,
while messages with at most $c=5$ recipients account for $97\%$;
the results here are stable across all these values of $c$,
and in this discussion we focus on the case of single-recipient messages.

Our second type of preprocessing is the follows.
Because not all members of the full population used their e-mail
addresses actively during this time, we focus on the $q$-fraction
of highest-volume e-mail users in this set, for various values of $q$.
In this discussion we use $q = .20$, defining
a set in which each user sent or received a message 
at least approximately once an hour during working hours 
for the full two-year time period.
However, the results discussed here are robust as
$q$ varies over a wide range.

\begin{figure}[t]
\begin{center}
\rs 
\includegraphics[height=\fht\textheight]{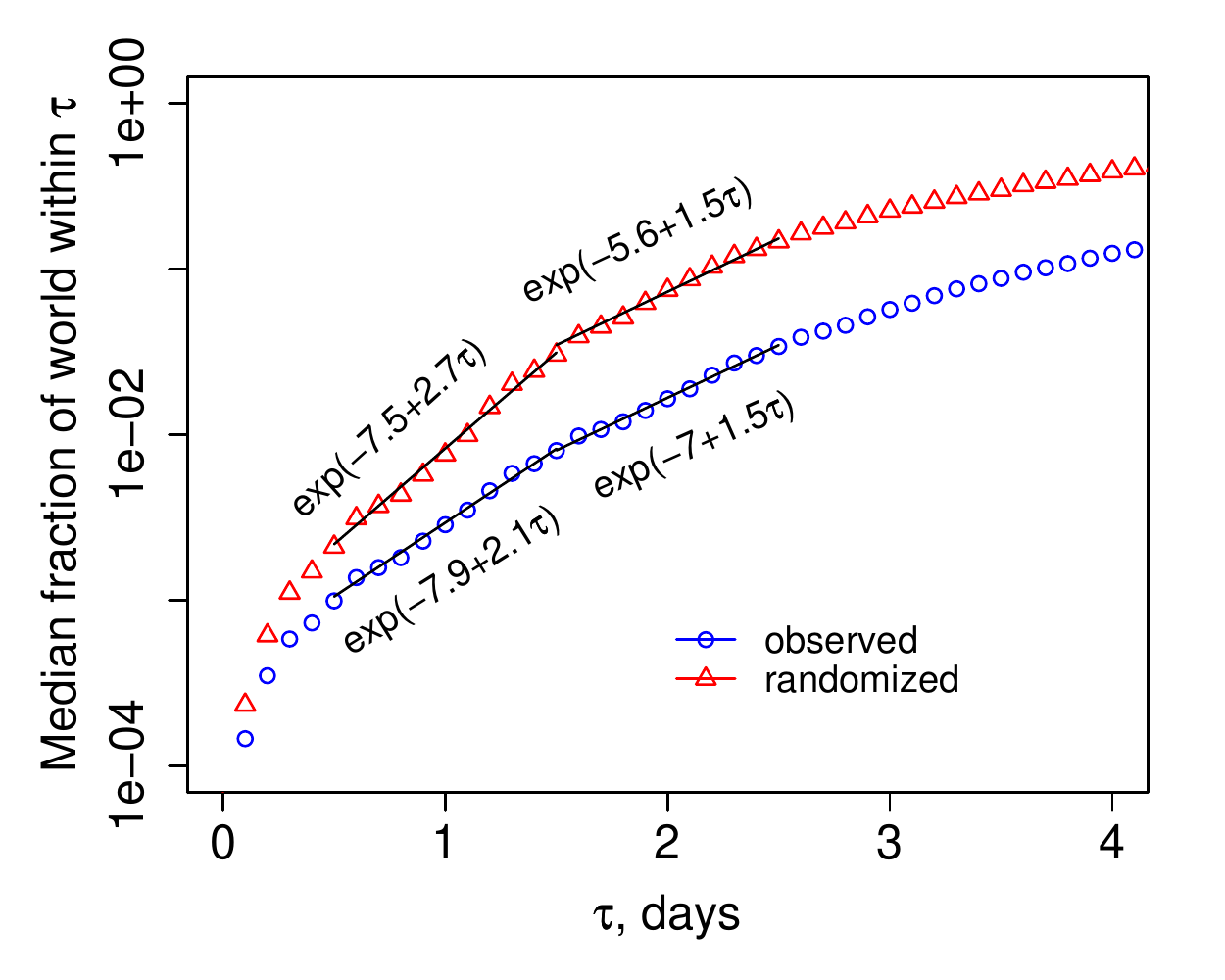}
\rs 
\caption{
The distribution of latencies among the $20\%$ of highest-volume
e-mail users.
\label{fig:latencies}
}
\rs \rs  
\halfrs
\end{center}
\end{figure}

We begin our analysis by considering
the distribution of information latencies ---
in other words, measuring how far
out-of-date the rest of the world is with respect to 
different nodes.
For a time difference $\tau$, we define the {\em ball of radius $\tau$
around node $v$ at time $t$}, denoted $B_{\tau}(v,t)$, to be the set
of all nodes whose latency with respect to $v$ at time $t$ is $\leq \tau$ days.
Now, for fixed $t$, the distribution of ball-sizes over nodes
can be studied using a function $f_t(\tau)$, defined as the median value of
$|B_{\tau}(v,t)|$ over all $v$;
this is simply the number of people who are within $\tau$ days
out-of-date of a typical node.
In Figure~\ref{fig:latencies} the lower curve plots (on a log-linear scale)
the average value of $f_t$
over $21$ fixed values of $t$, equally
spaced around one week to account for weekly variation.
We see that after an initial 12-hour ramp-up, the 
the number of people at $\tau$ days latency from
a typical node grows in an approximately piecewise exponential fashion.
The effect is that for a typical person $v$ in this community,
there are only about $12$ other people 
who are within a day and a half out-of-date with respect to $v$,
while there are over $200$ people within four days.

Extending this curve until the ball-size is half the community, we find
that the median latency between node pairs is $7.5$ days.
Now, to put the quantity $7.5$ days in context, we can compare it to other
possible measures of ``distance'' in the network.
If we look at unweighted distances (i.e. simple ``hop-counts'') 
in the communication skeleton $\skel$, we find that
the median distance between nodes is $3$,
a very small number characteristic of the {\em small-world} properties
of such networks \cite{travers-swn,watts-strogatz}.
But the simple fact that people in this community are
``three degrees of separation'' apart cannot be directly
translated into statements about the potential for information flow,
since that requires the temporal data that forms
the basis for our definition of information latency.

With temporal data in hand, we see that latency depends both
on the variation in {\em who} people communicate with and also the
on the variation in {\em how frequently} they communicate.
We can thus put the observed quantities in perspective
by holding the frequency of communication fixed, and
studying how the latencies change as we vary 
the choice of communication partners.
In particular, we compare the observed information latencies
with the results of a randomized baseline, as follows.
Suppose that we simulate the sequence of e-mail exchanges, except
that for each communication event, we have the sender contact a 
uniformly {\em random} person rather than their true recipient in the data.
In this way, the potential for information flow occurs
at the speed of a random epidemic, 
rather than according to the actual trace of e-mail communication.
The randomized latencies are generally shorter than
the real latencies, and the upper curve in Figure~\ref{fig:latencies} 
plots the median ball-sizes for this baseline.

These ball-sizes also grow in a roughly piecewise exponential fashion,
and the median latency among node pairs under 
randomized communication is $4.6$ days.
Interestingly, the local exponential growth rates
of the real latencies and the randomized baseline 
are roughly the same after about $36$ hours;
it is the faster exponential ``head start'' within this
first $36$ hours that allows the randomized baseline
to spread so much more quickly.
Essentially, under the real communication pattern,
the typical person resides in a kind of temporal ``bubble''
at the 36-hour radius, in which they can only be aware
of information from about 12 other people.
With randomized communication, on the other hand, 
information breaks out quickly to many people;
the median ball-size at 36 hours is already 50 people.
This initial difference plays a significant role in the 
different ball-sizes multiple days later.

\begin{figure}[t]
\begin{center}
\rs 
\includegraphics[height=\fht\textheight]{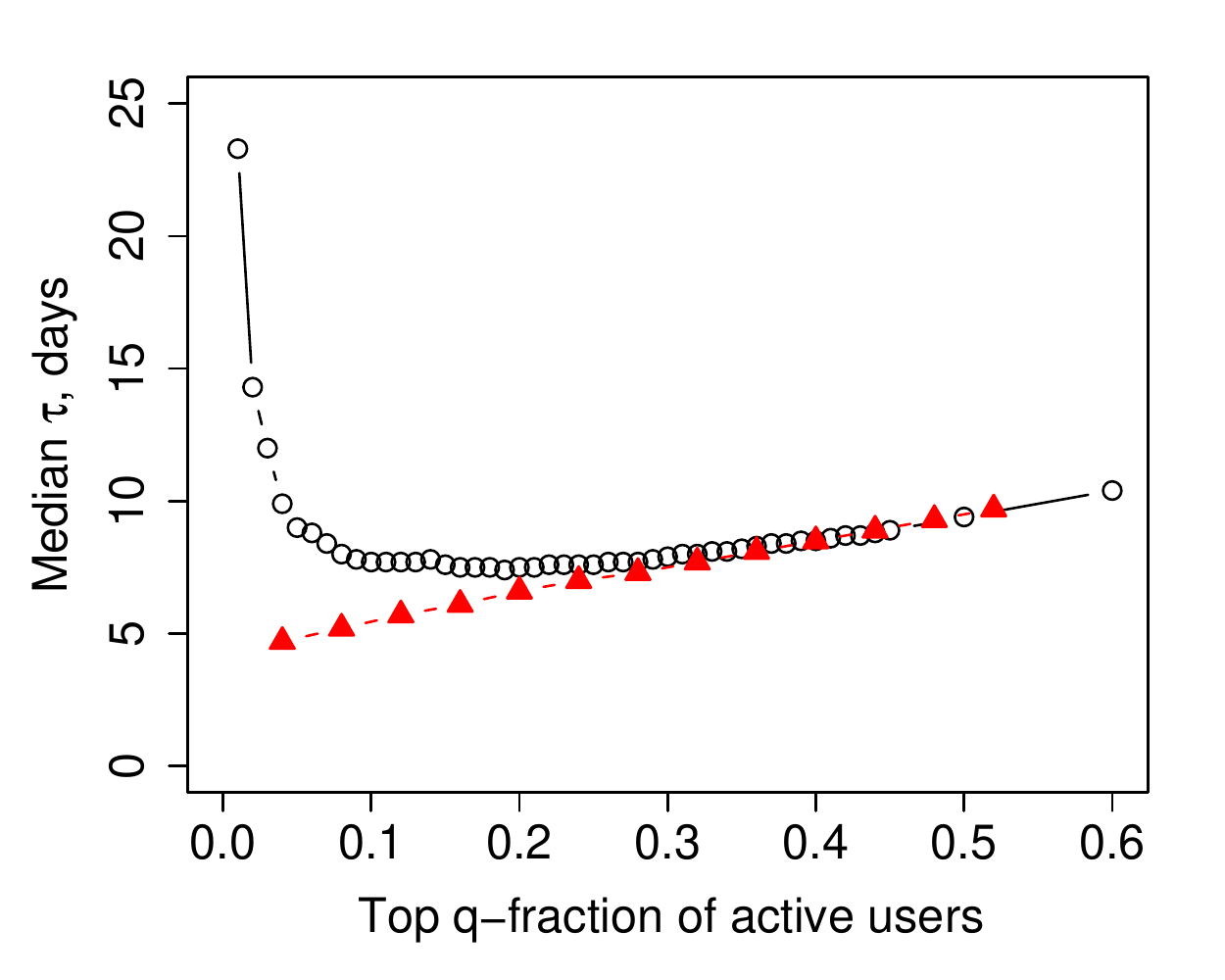}
\rs 
\caption{
The median latency for different $q$-fractions of the community,
both in isolation and embedded in the full community.
\label{fig:open-world}
}
\rs \rs  
\halfrs
\end{center}
\end{figure}

\xhdr{Open worlds vs. closed worlds}
Any dataset of communicating people $V$ will be typically embedded
in some much larger, unobserved set $V'$.
If we could watch the communication in this larger set $V'$, the
latencies even just among nodes in $V$ would decrease, due to 
quick paths between members of $V$ that snake in and out of $V' - V$.
We wish to understand this effect, so that we know how
to interpret latencies as we measure them in the ``closed world'' $V$
rather than the ``open world'' where $V$ is embedded in a larger $V'$.
In sociology, this is known as the {\em boundary specification problem}
\cite{kossinets-missing-data,laumann-boundary},
and it is inherent in essentially any study of a social
network embedded in some larger world.

We can address the effects of this issue in our context as follows.
Since we are studying the $q$-fraction of most active users in 
our university e-mail set, we can ask how median latencies
differ depending on whether we study this $q$-fraction in isolation,
or embedded in the full set of faculty and staff (the $q = 1.0$ fraction).
We show this in Figure~\ref{fig:open-world}: the upper curve 
plots median latency as a function of $q$ when the most active $q$-fraction
is observed on its own, and the lower curve plots the median latency
in the same set when it is observed embedded in the full community.
For extremely small values of $q$, the effect is considerable,
but once $q$ exceeds $0.1$, the effect becomes surprisingly negligible.

In addition to providing validation for the analysis of different
$q$-fractions in isolation, we believe this implicitly supports
a broader type of approximation --- specifically, when an active e-mail
network implicitly defines a natural community on its
members (as in the university community in this case),
it suggests ways to reason about it as a free-standing object despite the
fact that it is embedded in the unobservable global e-mail network.

\begin{figure}[t]
\begin{center}
\rs 
\includegraphics[height=\fht\textheight]{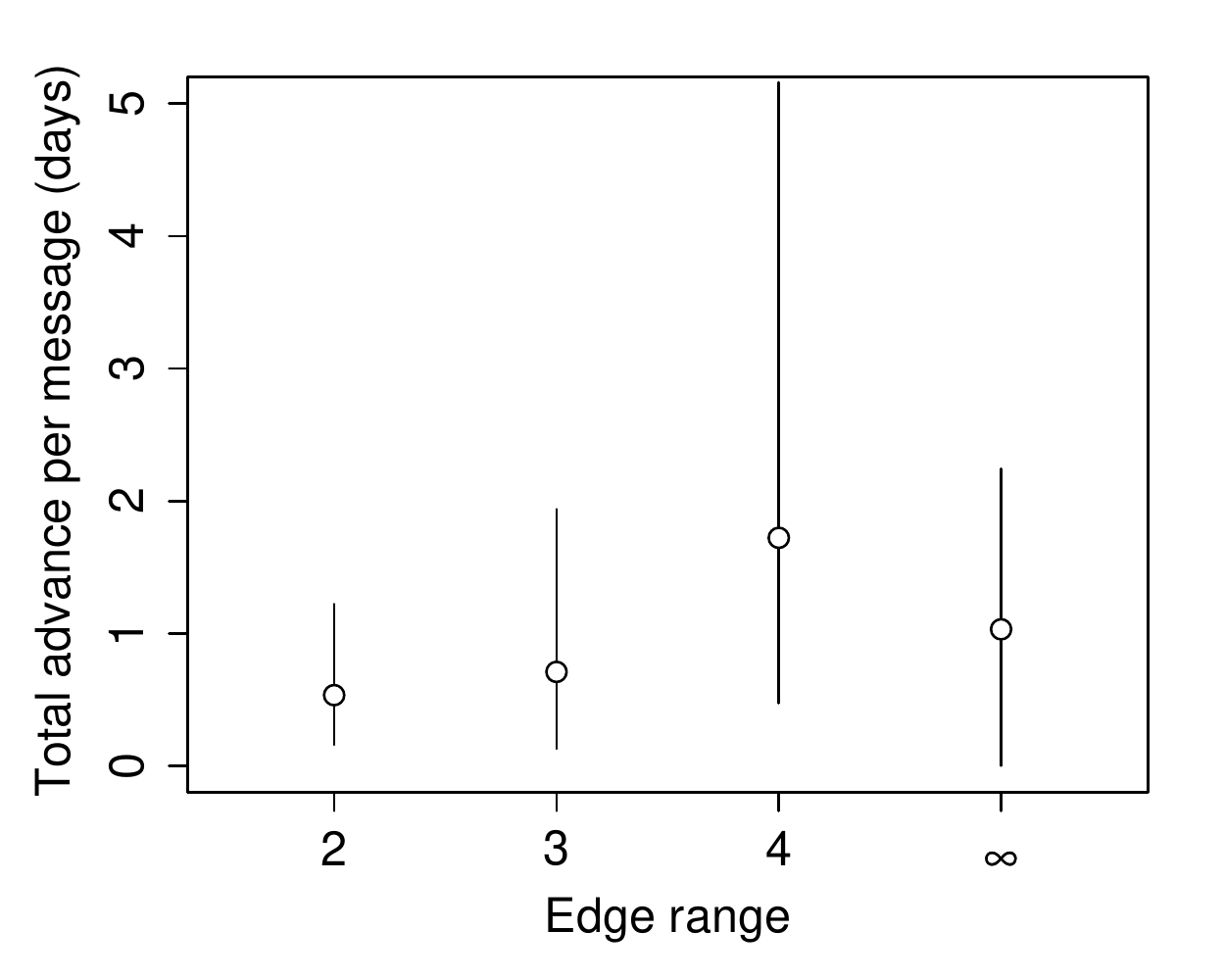}
\rs 
\caption{
The distribution of clock-advances per message as a function of
edge range.
\label{fig:weak-ties}
}
\rs \rs  
\halfrs
\end{center}
\end{figure}

\xhdr{Quantifying the strength of weak ties}
In a paper that has been very influential in sociology,
Granovetter proposed that 
{\em weak ties} --- connections to people
who form weaker acquaintance relationships, rather than close friendships ---
play an important role in conveying information to each of us
from parts of the social network that are inaccessible to our
circles of close friends \cite{granovetter-weak-ties}.
As a concrete example, Granovetter found that people very often reported 
receiving information leading to new jobs not from close friends,
but from more distant acquaintances; the close friends were perhaps
more motivated to help in tracking down job leads, but the more
useful information came through the distant acquaintances.

Granovetter formalized this by introducing a parameter
that we call the {\em range} of
an edge $e = (v,w)$, defined to be the unweighted shortest-path distance
in the social network between $v$ and $w$ if $e$ were deleted;
the range is thus the (unweighted) length of the
shortest ``alternate path'' between the endpoints
\cite{granovetter-weak-ties,watts-swn-book}.
Most edges in a typical social network will have range two,
indicating that $v$ and $w$ have at least one friend in common.
Granovetter's argument was that edges of range greater than two are 
generally weak ties --- i.e., edges connecting 
us to acquaintances with whom we have less frequent communication ---
and that these long-range edges are the sources
of important information to their endpoints.
However, he noted \cite{granovetter-swt-revisited} that 
despite interview-based methods to explore this principle,
it has been an open question to provide quantitative evidence
for it on social-network datasets.

We argue here that our vector-clock analysis can provide 
evidence for this phenomenon.
If we recall the algorithm that computes the vector clocks,
the basic step is to update the clock of a node $w$ when it
receives a message from some other node $v$.
Let us define the {\em advance} in $w$'s clock to be
the sum of coordinatewise differences between $\view_w$
before the update from $v$ and $\view_w$ after the update from $v$.
Intuitively, the advance is then the potential for new information
about the rest of the world that $w$ has gained as a result
of this single communication with $v$ --- a way of formalizing
the type of information-flow that Granovetter's work addresses.
To get at his observation, we can thus ask: if $(v,w)$ is an edge
in the communication skeleton $\skel$ 
of range greater than two, does each communication from $v$ 
result in an unusually large advance to $w$'s clock?

While this is a subtle effect to capture, we see evidence
for precisely this in Figure~\ref{fig:weak-ties}.
As a function of edge range $r$, we plot the median clock-advance per message
over all edges in $\skel$
at the given range $r$ (the open circles in the plot),
as well as the $25^{\rm th}$ and $75^{\rm th}$ percentiles
(the vertical line segments).
Due to the active communication within this group over two years,
there are no edges of finite range larger than four;
the infinite-range edges are bridges 
whose removal would disconnect the network.
(Since one side of each of these bridges is typically an extremely
small set of nodes, it is not necessarily surprising to see 
a typical clock advance that is smaller than the case of range $4$.)
In summary, we see that the clock-advance per message increases
with edge range, particularly for edges of range $4$,
thus suggesting that long-range
bridges can indeed be effective in transferring information from otherwise
distant parts of the network.

\section{Backbone Structures}

Having considered methods for analyzing the notion of out-of-date
information, we now use this to study the second issue mentioned
at the outset --- the structure of fast indirect paths ---
by introducing a concept that we call the {\em backbone}.

\xhdr{Defining the backbone}
To develop this idea, we start by recalling the observation
from the example in Figure~\ref{fig:timestamps}, where the direct $A$-$B$ edge
was a slower conduit for potential information from $A$ to $B$
than the indirect path $A$-$C$-$B$.
Let us say that an edge $(v,w)$ in the communication skeleton $\skel$ is
{\em essential} at time $t$
if the value $\view_{w,t}(v)$ is the result of a vector-clock
update directly from $v$, via some communication event $(v,w,t')$
where $t' \leq t$.
In other words, the edge is essential if $w$'s most up-to-date
view of $v$ is the result of direct communication from $v$,
rather than a sequence of updates along an indirect path from $v$ to $w$.
Thus for example, in Figure~\ref{fig:timestamps}, consider
all edges linking to $B$ in the communication skeleton: 
the edges $(C,B)$ and $(E,B)$
are essential at Friday 5pm, but the edges $(A,B)$ and $(D,B)$ are not.

We define the {\em backbone} $\bb_t$ at time $t$ to be the graph
on $V$ whose 
edge set is the collection of edges from $\skel$ that are
essential at time $t$.
(Although $\bb_t$ is a directed graph, we will also sometimes study 
properties of it as an undirected graph, simply by suppressing
the directions of the edges.)
Thus the backbone reflects
those communications responsible for all nodes' up-to-date views
at a given time $t$ --- i.e., those that are
not ``bypassed'' by some indirect path.

\begin{figure}[t]
\begin{center}
\rs \rs
\vspace{1ex} 
\includegraphics[height=0.22\textheight]{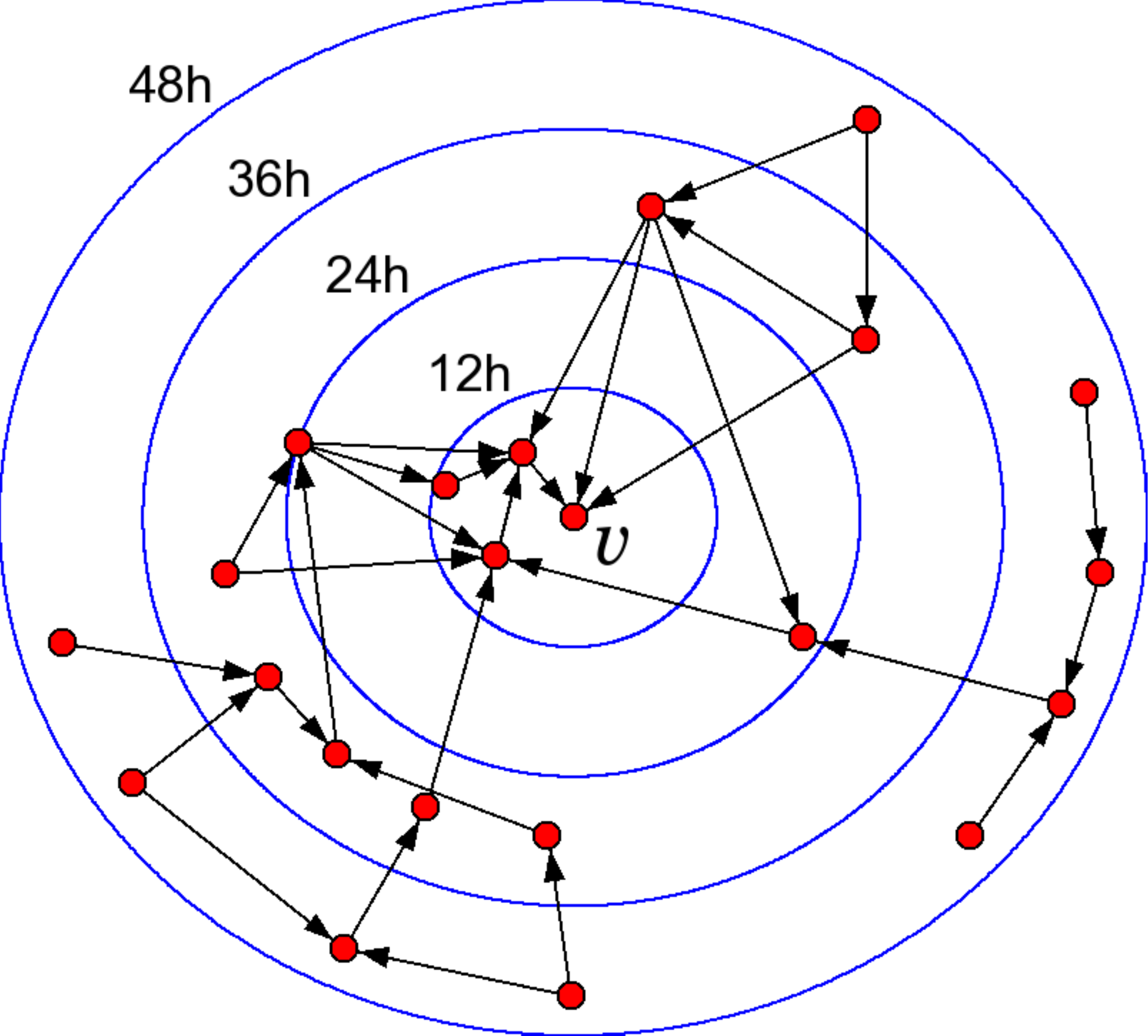}
\rs 
\caption{
A drawing of a small part of a backbone $\bb_t$ computed
from the university e-mail data, showing
only the portions induced on a particular
node $v$ and all nodes within 48 hours latency of $v$.  
Concentric circles denote ball radii increasing by 12 hours each,
and the distance of each node from the common center is its
latency from $v$.
\label{fig:vis}
}
\rs \rs  
\halfrs
\end{center}
\end{figure}

As a visual illustration,
Figure~\ref{fig:vis} depicts a small part of 
a backbone $\bb_t$ computed from the university e-mail data,
drawing only the portions induced on a particular node $v$
and all nodes within 48 hours latency of $v$.

\xhdr{An aggregate backbone}
The backbone is defined at each point in time via vector-clocks;
but it is also useful to have a single graph that summarizes 
in an analogous but simpler way 
the ``aggregate'' communication over the full two-year period,
and to be able to compare this simplified structure 
to the backbones defined thus far.
We can define such an aggregate structure by approximating 
communication between pairs of nodes as perfectly periodic.
For each edge $(v,w)$ in the communication skeleton
$\skel$ such that $v$ has sent $\rate_{v,w} > 0$
messages to $w$ over the full time interval $[0,T]$, 
we define the {\em delay} $\delay_{v,w}$ of the edge $(v,w)$ to 
be $T / \rate_{v,w}$.
This can be viewed as the gap in time between messages from $v$
to $w$, if communication from $v$ to $w$ were evenly-spaced.

Now consider the weighted graph $\wgr$ obtained from the
communication skeleton $\skel$ by assigning a weight of 
$\delay_{v,w}$ to each edge $(v,w)$.
The path of minimum total delay between two nodes $x$ and $y$ 
--- i.e., the path with minimum sum of delays on its edges ---
represents the fastest that information could flow 
from $x$ to $y$ in this ``aggregate'' setting where 
communication is evenly spaced.
We can now ask which edges are essential in an aggregate sense:
if, over the full time period studied, they are not bypassed 
by faster indirect paths.
Thus, we say that an edge $e = (v,w)$ in $\wgr$ is {\em essential} if
it forms the minimum-delay path between its two endpoints,
and we define the {\em aggregate backbone} $\abb$ to be the
subgraph of $\wgr$ consisting only of essential edges.
(For the sake of easier terminology, we will sometimes refer
to the backbones $\bb_t$ at fixed times $t$ as
{\em instantaneous backbones}, by contrast with the
aggregate backbone which is based on an aggregate 
construction that takes all times into account.)

We note that the construction of the aggregate backbone 
$\abb$ can be done more efficiently
than by simply considering each edge of $\wgr$ separately.
Rather, we can compute a weighted shortest-paths tree rooted
at each node of $\wgr$, using the delays as weights; 
the union of the edges in all these trees will be $\abb$, 
by the following proposition.

\begin{proposition}
An edge $e = (v,w)$ belongs to $\abb$ if and
only if it lies on the minimum-delay path between 
some pair of nodes $x$ and $y$.
\end{proposition}
\proof{
The ``only if'' direction is immediate, so we focus on
proving that if $e$ lies on the minimum-delay path 
between some pair of nodes $x$ and $y$, then it is essential.
We do this by contradiction: suppose $e$ lies on the 
minimum-delay path $P_{x,y}$ between some nodes $x$ and $y$, 
but it is not the minimum-delay path between its endpoints $v$ and $w$.
This means that there is a path $P_{v,w}$ of strictly smaller
delay than the edge $(v,w)$.
Now let $P_{x,v}$ be the subpath of $P_{x,y}$ from $x$ to $v$,
and let $P_{w,y}$ be the subpath of $P_{x,y}$ from $w$ to $y$.
Concatenating $P_{x,v}$, $P_{v,w}$, and $P_{w,y}$ would give
an $x$-$y$ path of strictly smaller delay than that of $P_{x,y}$,
which is the contradiction we seek. \rule{2mm}{2mm}
}

\begin{figure}[t]
\begin{center}
\rs 
\includegraphics[height=\fht\textheight]{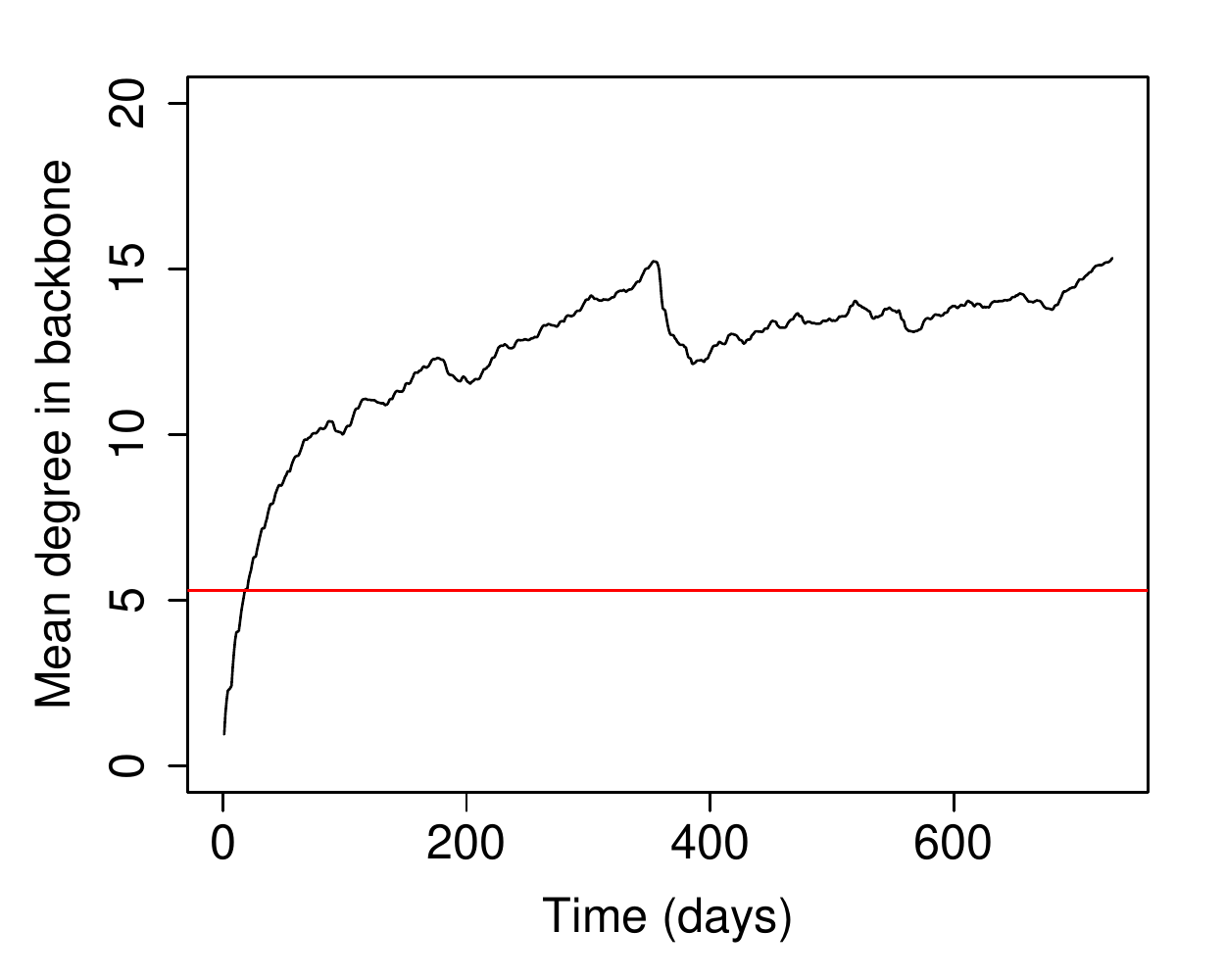}
\rs 
\caption{
The average degree over time in the backbone, with the
horizontal line depicting the average degree of $\approx 5$
in the aggregate backbone $\abb$.
\label{fig:density}
}
\rs \rs  
\halfrs
\end{center}
\end{figure}

\begin{figure}[t]
\begin{center}
\includegraphics[height=\fht\textheight]{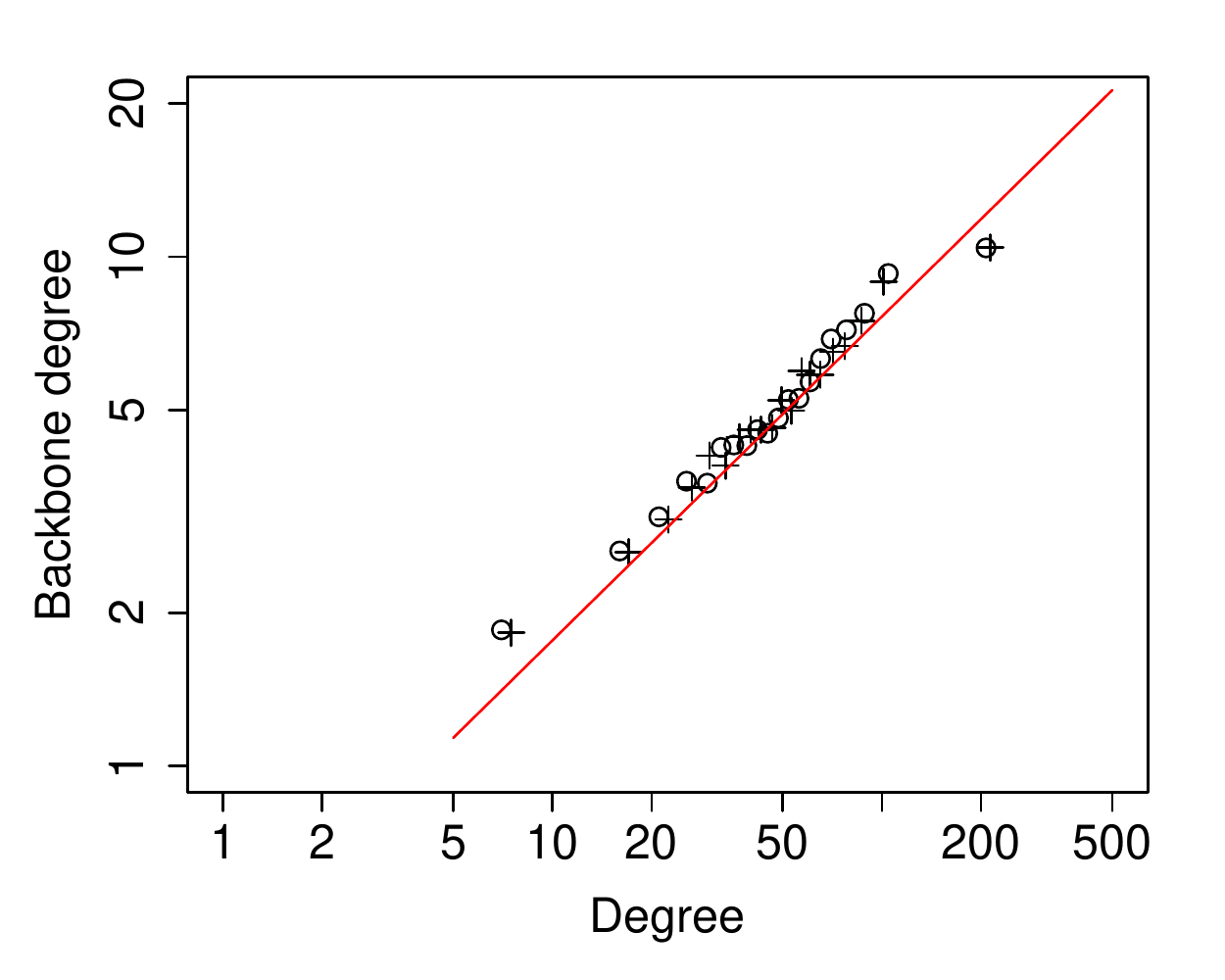}
\rs 
\caption{
In-degree (circles) and out-degree (crosses) 
in the aggregate backbone as a function of degree in 
the full communication skeleton $\skel$.
The sublinear growth indicates that 
the backbone eliminates edges from high-degree nodes 
at a greater rate.
\label{fig:degree-scatter}
}
\rs \rs  
\halfrs
\end{center}
\end{figure}

\xhdr{Density and node degrees of the backbones}
While the communication skeleton is a fairly dense graph,
we find that the backbones and the aggregate backbone are
surprisingly sparse --- in other words, from the point of
view of potential information flow, a significant majority of all edges
in the social network are bypassed by faster indirected paths.

In particular, 
Figure~\ref{fig:density} shows the
average degree in the instantaneous backbones $\bb_t$ as a function of time.
Note that there are clear boundary effects as the vector
clocks get ``up to speed,'' but after this initial phase
the average degree stabilizes to approximately $13$ even
as the backbone itself changes over time.
The aggregate backbone $\abb$ is sparser still: its
average degree is approximately five (the horizontal
line in Figure~\ref{fig:density}).
For comparison, the average node degree in 
the communication skeleton is approximately $50$.
In summary, even in this community of 
active users of e-mail, the typical person has only five
contacts that are not bypassed by shorter paths
in steady-state over a long time period.

The fact that the instantaneous backbones $\bb_t$ are roughly $2.5$ times
as dense as the aggregate backbone indicates the local
burstiness of communication in the network: at any particular
point in time, people have essential communication with
certain contacts that are not sustained in steady-state
over the full two-year interval.
It thus becomes natural to ask how much overlap there is
between the instantaneous backbones $\bb_t$ and the sparser aggregate backbone.
We find in fact that the overlap is substantial: each backbone $\bb_t$,
on average, contains roughly $3/4$ of the edges from $\abb$.
Of course, which {\em particular} edges of $\abb$ appear 
in any one $\bb_t$ varies considerably with $t$.
Thus, it is reasonable to think of the instantaneous backbones $\bb_t$
as roughly consisting of a large but varying piece of the aggregate backbone,
supplemented with transient edges whose membership in the
backbone changes more rapidly over time.

Considering the backbone also sheds further light on the role
of high-degree nodes in the social network.
It has been argued that high-degree nodes play 
a crucial function in the structure of
short paths in unweighted graphs \cite{albert-scale-free-deletion}.
It has also been argued, however, that the importance of these ``hubs''
diminishes considerably once temporal effects
are taken into account \cite{gibson-scheduling}. We find
support for both arguments: high-degree nodes in the full communication
skeleton $\skel$ indeed have many incident edges in the
aggregate backbone; however the fraction of a node's edges that are
declared essential strictly {\em decreases} with degree.
As Figure~\ref{fig:degree-scatter} illustrates, 
nodes of degree $k$ in $\skel$ have an average degree
of approximately $k^{0.6}$ in the aggregate backbone $\abb$;
thus, the fraction of a node's edges that are essential
is decreasing in its degree as $k^{-0.4}$.
A corresponding effect holds for the instantaneous backbones
$\bb_t$, where nodes of degree $k$ in $\skel$ have
average degree approximately $k^{0.65}$, an exponent
that remains stable over time after an initial start-up period.
Thus the backbones have a kind of ``leveling'' 
effect on the degrees, in which the spread between low
and high degrees is contracted faster than just proportionally
when we move from $\skel$ to its backbones.

\begin{figure}[t]
\begin{center}
\rs 
\includegraphics[height=\fht\textheight]{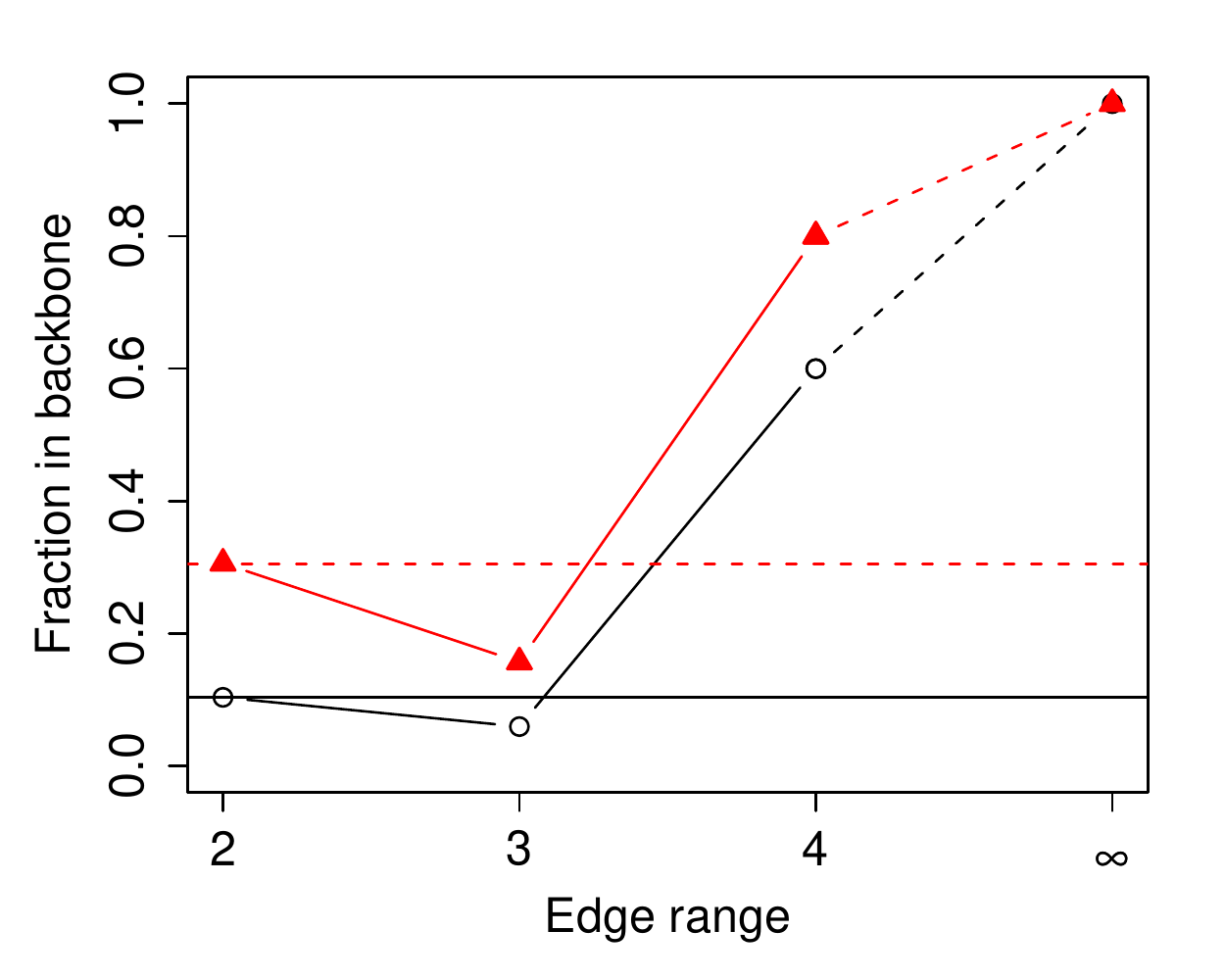}
\rs 
\caption{
Proportion of edges in the backbone for each edge range.
The lower curve is for the aggregate backbone and the
upper curve is for the instantaneous backbones.
The horizontal lines
represent the overall fraction of edges in the respective backbones.
\label{fig:range}
}
\rs \rs  
\halfrs
\end{center}
\end{figure}

\begin{figure}[t]
\begin{center}
\includegraphics[height=\fht\textheight]{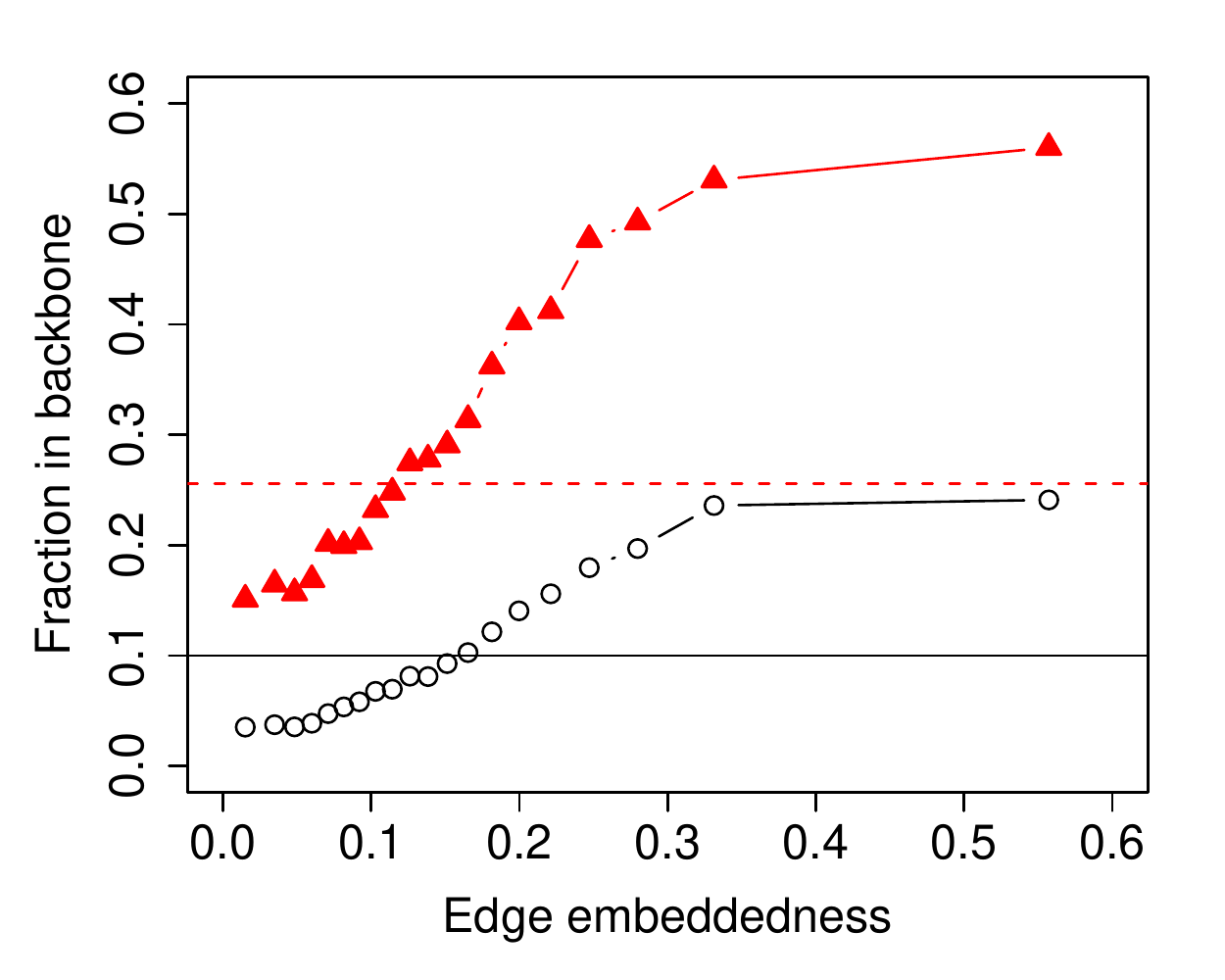}
\rs 
\caption{
Proportion of range-2 edges in the backbone as a function
of embeddedness.
Symbols are as in Figure~\ref{fig:range}.
\label{fig:embed}
}
\rs \rs  
\halfrs
\end{center}
\end{figure}

\xhdr{Structure of the backbone}
Intuitively, the backbone is trying to balance two competing
objectives: representing edges that span different parts of
the network, which transmit information at long ranges;
and representing very rapidly communicating edges,
which will typically be embedded in denser clusters and
transmit information at short ranges over quick time scales.
In fact, we will see in 
Figures~\ref{fig:range} and \ref{fig:embed} that
the mixture of edges in the backbone achieves precisely
a version of this trade-off.
For this discussion, we view the backbones as undirected
graphs simply by suppressing the directions of the edges.

In Figure~\ref{fig:range}, we show the proportion of edges
from $\skel$ that belong to the backbones, as a function of
their range.  (Recall that the {\em range} of an edge $e$
is defined as the distance between the endpoints of $e$,
when $e$ itself is deleted.) 
The lower curve depicts the aggregate backbone, while the
upper curve depicts the average over instantaneous backbones.
In each case, we see that there is an underrepresentation of edges of the
intermediate range $3$, with a greater density at the
two extremes of range $2$ and range $4$.
The large proportion of range-$4$ edges in the backbone
is another reflection of the strength-of-weak-ties
principle discussed earlier --- long-range edges
serve as important conduits for information.
To understand the picture at the other extreme, 
with edges of range $2$,
it is useful to further refine this set of edges 
using the notion of embeddedness.

We define the {\em embeddedness} of an edge to be,
roughly, the fraction of its endpoints' neighbors
that are common to both.
Formally, for an edge $e = (v,w)$, let $N_v$ and $N_w$
denote the sets of neighbors of the endpoints $v$ and $w$ respectively.
We define the embeddedness of $e$ to be
$|N_v \cap N_w| / |N_v \cup N_w|$.
Thus, highly embedded edges intuitively occupy dense clusters,
in that their endpoints have many neighbors in common.
We see in Figure~\ref{fig:embed} that highly-embedded edges 
are also overrepresented in both the aggregate and instantaneous backbones.
This may be initially surprising, since edges of large embeddedness
have many possible two-step paths that could
short-cut around them; their presence in the backbone
is thus a reflection of the generally elevated rate of communication
that takes places on such edges.

Taken together, then, these results on range and embeddedness
indicate a striking sense in which the backbone balances 
between two qualitatively different kinds of information flow:
flows that arrive at long range over weaker ties, and
flows that travel quickly through densely clustered regions in the network.

\section{Varying Speed of Communication}

We note that although the social communication patterns we are studying
arise organically (rather than being centrally designed), 
one can nevertheless study how the resulting latencies depend on
local variations in communication styles.
One could ask this question in the context of communication
within a large organization, for example: how do individuals'
decisions about communication strategies affect
the overall rate of potential information flow in the organization?
Of course, analysis of such questions can also potentially provide
insight into the design of information-spreading mechanisms
in engineered networks as well \cite{demers-epidemic}.

In particular, 
we study what happens to information latencies when each node
keeps its set of contacts the same, but 
varies the relative rates of its communication with these contacts.
Suppose we assume the communication skeleton $\skel$
represents the complete set of potential communication partners
for each person, and we allow people to change the individual rates
at which they send messages to these partners, while keeping their
total daily volume fixed.
Are there simple ways to change individual rates 
that will reduce the shortest-path delays
among pairs in the aggregate backbone?

As a baseline for comparison, we can consider the optimal reduction in delay,
given a central planner with complete knowledge of the network.
Here is a concrete way to formalize this optimization question in general.
We are given a directed graph $G$, with a {\em total rate}
$\rate_v$ for each node $v$.  We are also 
given a set $S$ of pairs of nodes in $G$ whose shortest-path
delays we want to reduce.
Each node $v$ can choose a rate $\rate_{v,w}$ at which to communicate
to each of its neighbors $w$, subject to the constraint that
$\sum_w \rate_{v,w} = \rate_v$.
These rates define delays $\delay_{v,w} = T / \rate_{v,w}$ as
in our construction of the aggregate backbone.
(Rates $\rate_{v,w}$ can be set to zero,
in which case the resulting edge $(v,w)$ is taken to have 
infinite delay.)
Now, the question is: for a given bound $\delay$, can
we choose rates for each node so that the median
shortest-path delay between pairs in $S$ in the aggregate
backbone is at most $\delay$?

As formulated, this optimization problem is intractable.

\begin{theorem}
The delay minimization problem defined above is NP-complete.
\end{theorem}
\sketch{We reduce from the 3-SAT problem.
Given a set of variables $x_1, ..., x_n$
and clauses to satisfy, we construct a
graph $G$, pairs $S$, and node rates $\rate_v$ as follows.
For each variable $x_i$, we construct three nodes $u_i, v_i, w_i$
with edges $(u_i,v_i)$ and $(u_i,w_i)$.
For each clause $C_j$, we construct nodes $s_j$ and $t_j$;
then, for each variable $x_i$ in clause $C_j$, we add edges
$(s_j,u_i)$ and $(v_j,t_j)$ if $x_i$ occurs positively in $C_j$,
and we add edges
$(s_j,u_i)$ and $(w_j,t_j)$ if $x_i$ occurs negatively in $C_j$.
Each node is given a rate of $1$.
Finally, we evaluate the median shortest-path delay
for the pairs $S = \{(s_j,t_j)\}$.

Now, if there is a satisfying truth assignment, then we can 
put a rate of $1$ on edge $(u_i,v_i)$ if $x_i$ is set to {\em True},
and a rate of $1$ on edge $(u_i,w_i)$ if $x_i$ is set to {\em False}.
We can also put a rate of $1$ on edges $(s_j,u_i)$ where
$x_i$ is a variable that satisfies $C_j$.
This makes all shortest-path delays between pairs in $S$ equal to $3$;
conversely, if the median shortest-path delay between pairs in $S$ 
can be made equal to $3$, then each pair in $S$ must have delay 
$3$, in which case a satisfying assignment can be determined
from the paths that are used.}

\begin{figure}[t]
\begin{center}
\rs 
\includegraphics[height=\fht\textheight]{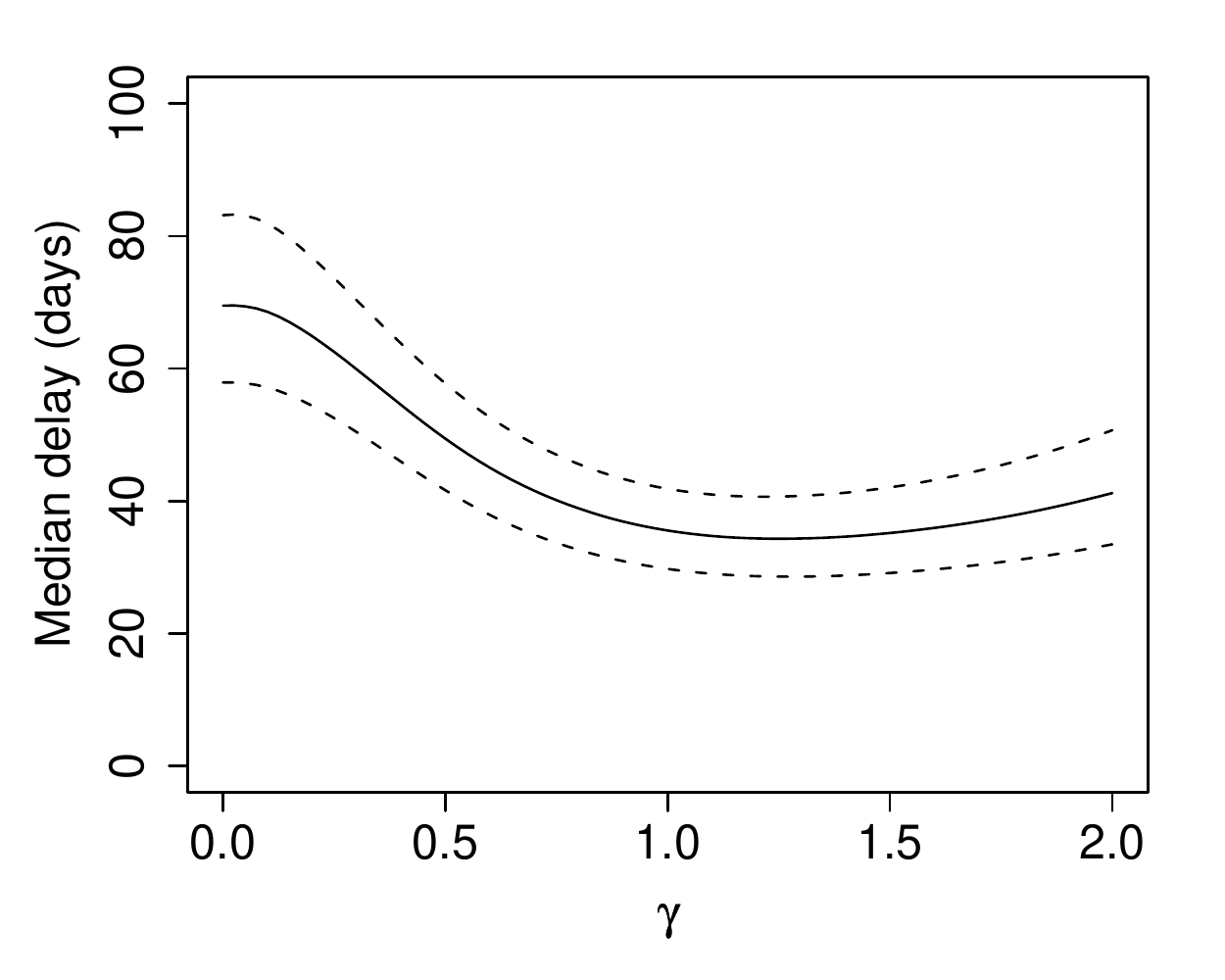}
\rs 
\caption{
Median shortest-path delay in the aggregate backbone (solid line)
as a function of the load-reweighting parameter $\gamma$.
Dashed lines represent the $25^{\rm th}$ and $75^{\rm th}$
percentiles of the shortest-path delay.
\label{fig:gamma}
}
\rs \rs  
\halfrs
\end{center}
\end{figure}

\xhdr{Load-leveling vs. load-concentrating}
While this intractability shows
the difficulty in optimally accelerating communication, 
a more realistic goal is to consider simple {\em local rules}
by which individuals in a network might vary their rates
of communication so as to influence the potential for information flow.
A basic qualitative version of this question is the following:
for accelerating potential information flow,
is it better to talk even more actively to one's most frequent
contacts, or to balance things out by increasing communication
with the less frequent contacts?
We could refer to the former strategy as 
{\em load-concentrating}, since it pushes more traffic
onto the already-high-volume edges, and we could refer
to the latter strategy as {\em load-leveling},
since it tries to level out the traffic across edges.

We can study this in the university 
e-mail data by choosing a re-scaling exponent
$\gamma$ and modifying the rates of communication on the
edges emanating from each node $v$, changing $\rho_{v,w}$
to $\rho_{v,w}^\gamma$ and then normalizing all rates from
$v$ to keep its total outgoing message volume the same.
Varying $\gamma$ thus smoothly parametrizes a family of
different strategies,
with values $\gamma > 1$ corresponding to load-concentrating
strategies --- since already-large rates are amplified ---
while values $\gamma < 1$ correspond to load-leveling strategies.

In Figure~\ref{fig:gamma}, we show the effect of these 
strategies on the median shortest-path delay in the
aggregate backbone.  We note, first of all, the interesting
fact that $\gamma = 1$ is close to the best possible for shortest-path delays;
in other words, the existing rates of communication
are close to optimal, in terms of potential information flow,
over this class of strategies.
However, there is still room for improvement in the shortest-path delay:
the optimal median, over all $\gamma$, occurs at $\gamma^* \approx 1.2$.
The fact that $\gamma^* > 1$ indicates an interesting and
perhaps unexpected result: that increasing the rate of communication
to the most frequent contacts actually has the effect of 
reducing shortest-path delays --- a result at odds with
the intuition that making stronger use of infrequent contacts
and weak ties is the way to reduce latency.

\xhdr{Node-dependent delays}
There is an extension of the model that sheds further light on this finding.
Suppose we extend the notion of delay to have not just
delays $\delta_{v,w}$ on each edge, but also a fixed delay
of $\varepsilon$ at each node, so that the total delay on
a path becomes the sum of the edge {\em and} node delays.
In other words, as information flows 
it incurs additional delays from each node that handles it.

Naturally, as $\varepsilon$ increases, there is a larger
penalty for paths that take more hops, and minimum-delay
paths increasingly come to resemble those of minimum hop-count.
This leads to a denser backbone, as fewer edges are
rendered inessential.
The value of $\gamma$ at which
network latency is optimized decreases with $\varepsilon$, 
crossing $\gamma^*=1$ at $\varepsilon \approx 4$
days. Thus, as the speed of diffusion pathways is determined
increasingly by node-specific (rather than edge-specific) delays, the
backbone becomes denser, and the importance of
quick indirect paths diminishes. Moreover, as node delays increase,
the optimal re-scaling of communication 
for reducing network latency transitions from load-concentrating to
load-leveling.

\section{Conclusions}

The basic definitions of social network analysis have
been primarily built on graph-theoretic foundations
rooted in unweighted graphs.
Here we have explored how this perspective changes when
one makes integral use of information about
how nodes communicate over time.
Rather than explicitly tracking the content of this
communication, we develop structural measures based
on the {\em potential} for information to flow;
in this way, we can get at elusive notions around
the network's everyday rhythms of communication.

With this view, some of the direct connections 
in the network become much longer,
due to low rates of communication, while other multi-step
paths become much shorter, due to the rapidity with
which information can flow along them.
We find that adapting the notion of {\em vector-clocks}
from the analysis of distributed systems provides a
principled way to measure how ``out-of-date'' one person
is with respect to another, and we find that the 
sparse subgraph of edges most essential to keeping people
up-to-date --- the {\em backbone} of the network ---
provides important structural insights that relate
to embeddedness, the role of hubs, and the strength of weak ties.
Finally, this style of analysis allows us to study
the effects on information flow as nodes vary the rate
at which they communicate with others in the network,
ranging from strategies in which communication is concentrated
on heavily-used edges to those in which it is leveled out
across many edges.

This style of analysis is applicable to any setting in which
a group of individuals is engaged in active communication
with the goal of exchanging information, and when there is
data available on the temporal sequence of communication events.
As discussed earlier, we have also explored the measures
defined here in other e-mail datasets (the Enron corpus),
as well as in settings that are quite different from e-mail networks ---
in particular, we have applied vector-clock and backbone
analysis to the communications among admins and other high-activity
editors on Wikipedia, using edits to user-talk pages as communication events.
Wikipedia is a setting where it is particularly easy to get public data
with complete communication histories, but it is also representative
of communities that maintain themselves through on-line communication
and coordination (large open-source projects and large media-sharing
sites are other examples).

Although the dynamics and patterns of communication in all
three of our datasets are quite different,
we find that a large number of the qualitative findings
discussed for the university e-mail domain 
carry over to the other settings studied, including
the sparsity of the aggregate and instantaneous backbones
and the variation in node degrees.
In particular, the typical aggregate backbone degrees around $5$ and the
recurring sub-linear ``compression''
of degrees --- where nodes of degree $k$ in the full skeleton
have typical backbone degree $\sim k^c$ for $c \approx 0.5$--$0.6$ in all three 
datasets --- are common patterns that seem to call for 
a deeper theoretical explanation.
On the other hand, compared to the university e-mail dataset,
we find that the ``core'' of active 
communicators is much smaller in both the Enron corpus 
(since it is data from a limited set of employees' mailbox folders)
and in Wikipedia (due to the specifics of community dynamics),
and this makes the {\em range} of an edge in the unweighted
communication skeleton harder to interpret and to correlate
with other measures for both these other domains.
In a sense, this is natural: principles about long-range edges
and their effects are derived from properties of large populations
with natural sub-communities --- as we find in the university e-mail data ---
and it is not clear that long-range edges carry the same
meaning in much smaller populations.

In general, we see the analysis framework proposed here 
as a way of comparing the different kinds of communication
dynamics within different communities.
Further investigation of these notions could ultimately shed light on
the principles that govern the
dynamics of different types of information, and how these principles
interact with the directed, weighted nature of social communication
networks.


\end{document}